\documentclass[aps,pra,manuscript]{revtex4}
\usepackage{graphicx}
\usepackage{epsfig}
\usepackage{color}
\usepackage{bm}

\begin{document}
\draft
\title{Geometry-diversified Coulomb excitations in trilayer AAB stacking graphene \\}
\author{Chiun-Yan Lin $^{1}$, Bor-Luen Huang $^{1}$, Godfrey Gumbs$^{2,3}$, Ming-Fa Lin$^{1,4,5}$}
\affiliation{$^{1}$Department of Physics, National Cheng Kung University, Taiwan\\
$^{2}$Department of Physics and Astronomy, Hunter College of the City University of New York, 695 Park Avenue, New York, New York 10065, USA\\
$^{3}$Donostia International Physics Center (DIPC), P de Manuel Lardizabal, 4, 20018 San Sebastian, Basque Country, Spain
$^{4}$Hierarchical Green-Energy Materials Research Center, National Cheng Kung University, Tainan, Taiwan\\
$^{5}$Quantum topology center, National Cheng Kung University, Tainan, Taiwan\\
}

\date{\today}

\begin{abstract}
The lower-symmetry trilayer AAB-stacked graphene exhibits  rich electronic properties and thus  diverse Coulomb excitations. Three pairs of unusual valence and conduction bands create nine available interband excitations for the undoped case, in which the imaginary (real) part of the polarizability shows 1D square root asymmetric peaks and 2D shoulder structures (pairs of antisymmetric peaks and logarithm type symmetric peaks). Moreover, the low frequency acoustic plasmon, being revealed as a prominent  peak in the energy loss spectrum, can survive in a narrow gap system with the large-density-of-states from the valence band. This type of plasmon mode is similar to that in a narrow gap carbon nanotube. However, the decisive mechanism governing  this plasmon is the intraband conduction state excitations. Its frequency, intensity and critical momentum exhibit a non-monotonic dependence on the Fermi energy. The well-defined electron-hole excitation boundaries and the higher frequency optical plasmons are transformed by varying  the Fermi energy. There remain substantial differences between the electronic properties of trilayer AAB, ABC, AAA and ABA graphene stackings.\end{abstract}

\vskip 0.2in


\medskip
\par
\maketitle

\par\noindent * Corresponding author. ~{{\it E-mail address}: ggumbs@hunter.cuny.edu (G. Gumbs)}

\par\noindent * Corresponding author. ~{{\it E-mail address}: mflin@mail.ncku.edu.tw (M. F. Lin)}

\section{Introduction}
\label{sec1}


\vskip 0.6 truecm

Experimental examination of the electronic properties
of low-dimensional materials has utilized a
variety of methods such as angle-resolved photoemission spectroscopy
(ARPES) \cite{hermanson,Ohta,Jin} and electron energy loss spectroscopy (EELS)
\cite{EE1,EE2,EE3,EE4,EE5}.
In a remarkable paper, Ohta, et al. \cite{Ohta} used ARPES to study
the screening of the  length,  and  strength of interlayer electronic  interaction in multilayer
graphene, in comparison  with  tight-binding calculations. This same
technique was also used by Jin, et al. \cite{Jin} to determine the thickness-dependence
of the electronic band structure of MoS$_2$.   The role played by EELS has been equally impressive. For example,
Pellegrino, et al. \cite{EE1} investigated the effect  due to strain on the optical conductivity of graphene. In Ref.\ [\onlinecite{EE2}], high-resolution EELS  was used to study the low energy collective excitations for Cu(111) surface. A plasmon mode with negative dispersion and an energy of 1.1eV was measured. It was suggested that this collective mode comes from the oscillation of electrons constrained to be in two-dimensional (2D) Shockley surface states  on the Cu(111) surface.
In the work of Lu, et al. \cite{EE5}, the plasmon dispersion was measured
for epitaxial graphene  usingh-resolution  EELS. There it was revealed that
the $\pi$ plasmons for single, bilayer and  trilayer graphene have
significant differences in their behavior.  The differences for  the in-phase
and out-of-phase modes arise from different band structures for single-layer and few-layer graphene.

\medskip
\par

These aforementioned facts have inspired us to use the tight-binding model to calculate the energy band structure and consequently investigate the behavior of the $\pi$ electron Coulomb excitations in AAB
trilayer graphene in the random-phase approximation (RPA). The dependence on the magnitude ($q$) and direction ($\phi$) of the transferred momentum and the Fermi energy ($E_F$) is carefully analyzed.

\medskip
\par

We note that  AAB stacking is a  combination of AA and AB stackings, as it is obviously apparent from
Fig. 1(a). Such a system has been successfully synthesized by  reliable methods and investigated, e.g., using an STM tip \cite{PRB86;085428}, mechanical exfoliation directly
by scalpering or scotch tape \cite{SurfSci601:498}, CVD growth on a SiC substrate\cite{PRB79:125411} and Ru(0001) surface \cite{APL107:263101}, and  liquid-phase exfoliation of natural graphite in N-methyl-2-pyrrolidone \cite{APL102:163111}.
In contrast,  ABA stacking could be obtained by rotating or horizontal shifting of the top graphite layer along the armchair direction. That is to say, the stacking configuration is continuously changed under electrostatic modulation of the STM. Additionally, the corresponding density-of-states (DOS) is also measured, indicating a narrow energy gap in trilayer ABA stacking \cite{APL107:263101,SciRep6:33487}.
According to the first-principles method, the ground state energies per unit cell of six
carbon atoms in trilayer graphenes are evaluated for our stacking configurations. They are
estimated as follows: $-$55.832866 eV, $-$55.857749 eV, $-$55.862386 eV and $-$55.864039 eV
for AAA, AAB, ABA and ABC stackings, respectively \cite{CRCPress;9781138556522}. The theoretical
calculations predict that the AAB stacking is more stable than the AAA one, or the former presents
the more promising future in experimental syntheses. The lower-symmetry AAB stacking possess the
most complicated interlayer hopping integrals, in which this special property is clearly identified
from a consistent/detailed comparison between the tight-binding model and VASP calculations for
the low-lying energy bands \cite{PCCP18:17597,Carbon94:619}. For example, the AAB-stacked trilayer
graphene possesses three pairs of energy bands with  oscillatory, sombrero-shaped and parabolic
dispersions, in which the first ones determine a very narrow band gap of ${<10}$ meV.
In particular, an acoustic plasmon mode in the long wavelength limit identifies the critical mechanism due to the large DOS of the oscillatory valence bands in the semiconducting pristine system or the free conduction electrons in the doping case.
We conduct the first theoretical study of the effects due to doping on the Coulomb excitations and we describe the related diversified phenomena in this work.

\medskip
\par

There exist some theoretical studies for AAB-stacked few-layer graphene structures on the band structure, density-of-states, magnetic quantization, and magneto-optical properties \cite{IOP2017,Carbon94:619,PCCP18:17597}. However, the significant research works dealing with  the many-particle carrier excitations and deexcitations are absent as far as we know. This complex possesses a very narrow band gap and unusual energy dispersions. Therefore, the electron-lectrone Coulomb interactions are expected to create rich and unusual excitation phenomena, such as the temperature- and doping-created low frequency plasmons. The former characteristics have been identified from  STS measurements.  It has also been observed that most pristine layered graphene exhibits semi-metallic behavior, e.g., the AAA, ABA and ABC stacking \cite{IOP2017} and the bilayer twisted \cite{PRB95;245116} and sliding systems \cite{SciRep4;7509}. From the tight-bindig model \cite{Carbon94:619} and the first-principles method \cite{CRCPress;9781138556522},  there exist three pairs of low-lying valence and conduction bands, with  oscillatory, sombrero-shaped and parabolic dispersions. They will exhibit the unusual optical vertical excitations, mainly owing to the unusual van Hove singularities in the DOS \cite{IOP2017}. Also, such energy bands induce  magnetically quantized Landau levels, with non-crossing, crossing and anti-crossing $B_z$-dependent energy spectra \cite{Carbon94:619}, being quite different from those of monolayer, bilayer and AAA/ABA/ABC trilayer graphene systems \cite{IOP2017}.  The lower-symmetry AAB stacking has been predicted to have more complicated magneto-optical absorption spectra \cite{IOP2017}.
\medskip
\par

The electronic properties and Coulomb excitations of the lower-symmetry trilayer AAB are, respectively, investigated with the use of the tight-binding model and the modified RPA in detail. Specifically, the intralayer as well as the interlayer hopping integrals and the intralayer and interlayer electron-electron interactions are taken into consideration simultaneously through the layer-dependent polarization functions. The low-lying energy bands mainly come from the ${2p_z}$ orbitals of  six carbon atoms in a primitive unit cell, being examined with the first-principles calculations \cite{CRCPress;9781138556522}. Three pairs of valence and conduction bands are identified to produce the unusual energy dispersions: the oscillatory, sombrero-shaped and parabolic ones, as measured from the Fermi level. The first two kinds of energy bands are predicted to yield   van Hove singularities in the density-of-states as  square root asymmetric peaks. That is, they could be viewed as  quasi-1D parabolic bands. Moreover, there exists a very narrow energy gap ($<$10 meV). The single-particle excitations are expected to exhibit a lot of special structures, since the nine categories of the interband transitions, being related to the low-symmetry wave functions, are effective under  the non-vertical Coulomb excitations. The current study clearly shows that a prominent plasmon peak, with  low frequency ($<$0.2 eV), appears in the energy loss function, and it is shown to originate from the interband excitations of the first pair of energy bands. The dependence on the transferred momentum and temperature will be explored thoroughly. Furthermore, the similarities between 2D AAB trilayer stacking and 1D narrow gap carbon nanotubes are  discussed in detail. The way in which one can substantially change the low frequency plasmon modes by  electron/hole doping is another focus in this work. The number,intensity, frequency and optical/acoustic mode of the  Fermi level related plasmons and the various regions of intraband and interband electron-hole excitations are worthy of a systematic investigation, the disparate (momentum, frequency)-phase diagrams with  specific plasmon modes and  distinct electron-hole boundaries. Finally, a detailed comparison among AAA, ABC, ABA and AAA stacking is made in the single- and many-particle properties.

\medskip
\par

\section{Energy band structure}
\label{sec2}

\medskip
\par

The low-lying energy bands of the trilayer AAB stacking, which principally originate from the ${2p_z}$ orbitals, were calculated with the tight-binding model. They are almost consistent with those from  the first-principles method
\cite{CRCPress;9781138556522}. The first two layers (the second and third layers), as clearly shown in Fig. 1, are arranged in the AA (AB) stacking. In this system, the A atoms (black spheres) possess the same ${(x,y)}$ coordinates, while the B atoms (red spheres) on the third layer are projected onto the hexagonal centers of the other two layers. The interlayer distance and the CVC bond length are, respectively, ${d=3.37}$ $\AA$ and ${b=1.42}$ $\AA$. There are six carbon atoms in a primitive unit cell, being similar to those in  trilayer AAA, ABA and ABC stackings. The low energy electronic properties are characterized by the complicated atomic interactions of the carbon ${2p_z}$ orbitals. The zero field Hamiltonian, being built from the six tight-binding functions associated with the periodical ${2p_z}$ orbitals, is dominated by the intralayer and the interlayer atomic interactions, ${\gamma_i}$'s. As a result of the lower symmetry stacking configuration, there exist ten types of  hopping integrals in the Hamiltonian matrix elements. ${\gamma_0\,=-}$2.569 eV represents the nearest-neighbor intralayer atomic interaction; ${\gamma_1\,=-0.263}$ eV $\&$  ${\gamma_2\,=}$0.32 eV, respectively, present the interlayer atomic interactions between the first and second layer;  ${\gamma_3\,=-}$0.413 eV/${\gamma_4\,=-}$0.177 eV/${\gamma_5\,=-}$0.319 eV are associated with the interayer atomic interactions between the second and third layer; ${\gamma_6\,=-}$0.013 eV, ${\gamma_7\,=-}$0.0177 eV, $\&$ ${\gamma_8\,=-}$0.0319 eV relate to the interlayer atomic interactions between the first and third layer; ${\gamma_9\,=-}$0.012 eV accounts for the difference between the chemical environments for A and B atoms. The specific hopping integrals, $\gamma_1$, $\gamma_3$ and $\gamma_5$ denote the vertical interlayer hopping integrals, while the others are non-vertical ones.  We also note that such tight-binding parameters are thoroughly examined to characterize the complex and unique energy bands of AAB-stacked trilayer graphene [Fig. 1(a)]. In addition, their magnitudes  are comparable with those used in the other stacking systems, e.g., AAA \cite{PRB46;4531}, ABA \cite{PRB44;13237} and ABA stackings \cite{PRB44;13237}.

\medskip
\par

The first, second, and third pairs of valence and conduction bands in trilayer AAB stacking, measured from the Fermi level, possess unique energy dispersions. These are the oscillatory (${S_1^{e.,v}}$), sombrero-shaped (${S_2^{c,v}}$) and parabolic bands, as clearly shown in Figs. 2(a) and 2(b). The first conduction band begins to increase from a local minimum value of ${\sim\,4}$ meV at the K point, along the K$\Gamma$ and KM directions. After reaching a local maximum value of ${\sim\,60}$ meV, it is decreased, returns to a local minimum energy again ($\sim$4 meV), and then increases steadily. The first conduction band is almost symmetric with the first valence band with respect to ${E_F=0}$, with corresponding  opposite curvature. As a result, there exist a narrow energy gap of ${E_{g}\sim\,8}$ meV and four constant-energy contours at ${\pm\,4}$ meV $\&\pm$ 60 meV. Specifically, such loops in  3D energy-wave vector space could be effectively regarded as  1D parabolic bands [Fig. 3(a)]. The multifold degenerate states are not suitable for doing the low energy expansion, i.e., the effective mass model cannot deal with the low energy essential properties. As for the ${S_2^{c,v}}$ energy bands, their conduction bband (valence band) states have a sombrero-shaped energy dispersion with local energy minimum (maximum) and maximum (minimum), being located near 0.255 eV ($-$0.255 eV) and 0.26 eV ($-$0.26 eV), respectively. The very small energy difference between two extrema points is ${\sim\,5}$ meV. Specifically, the third pair of ${S^{c,v}_3}$ consists of monotonic parabolic bands, for which the minimum (maximum) conduction (valence) state energy minimum is about 0.49 eV ($-$0.49  eV. The above-mentioned features of the low-lying energy bands are in agreement with those obtained from the first-principle calculations (the dashed curves in Figs. 2(a) and 2(b)), clearly indicating that the complicated interlayer hopping integrals employed in the tight-binding model are appropriate and reliable \cite{CRCPress;9781138556522}.  Such energy dispersions also produce the unusual magnetic quantization and magneto-optical properties
\cite{Carbon94:619,PCCP18:17597}, such as the frequent intragroup and intergroup LL anti-crossings, and the  intricate and rich magneto-optical absorption spectra,  being never revealed in other stacking systems.

\medskip
\par

The unusual energy dispersions generate two kinds of van Hove singularities in the density-of-states for trilayer AAB stacking. The asymmetric square root peaks and the shoulder structures, respectively, arise from the constant-energy contours (Fig. 2(a)) and the band-edge states of the parabolic dispersions. Two pairs of asymmetric peak structures (Taiwanese temple structures) which are related to the first pair of oscillatory energy bands, are centered about the Fermi level. The higher-energy/deeper-energy asymmetric peak only appears on the right-hand/left-hand side, since the sombrero-shaped conduction/valence band is too shallow. The rich density-of-states structures lead to very complex optical excitations, for which there are nine kinds of available vertical transitions arising from three pairs of valence and conduction bands. The number, frequency, intensity and form of the optical absorption structures are predicted to be very sensitive to the strength of an external electric field \cite{IOP2017}.

\section{Theory}
\label{sec3}

When an electron beam is incident on the AAB-stacked trilayer graphene, the charge density distribution is assumed to be uniform inside each layer. The $\pi$ electrons on specific layers will screen the time-dependent external electrostatic potentials (${V_{ll^\prime}({\bf q})}$'s; the $l$-th layer) via electron-electron interactions, leading to  induced charge density fluctuations and potentials. Within  linear response theory, the induced charge density is proportional to the effective Coulomb potentials (${V^{eff}_{ll^\prime}({\bf q},\omega\,)}$'s; $\omega$ is the transferred frequency during charge screening). By using the layer-based RPA, the relationship between the effective, external and induced Coulomb potentials is characterized by  Dyson's equation\cite{PRB74;085406}

\begin{equation}
\epsilon_{0}V^{eff}_{ll^{\prime}}(\mathbf{q},\omega)=V_{ll^{\prime}}(\mathbf{q})+
\sum\limits_{m,m^{\prime}}V_{lm}(\mathbf{q})P^{(1)}_{m,m^{\prime}}(\mathbf{q},\omega)
V^{eff}_{m^{\prime}l^{\prime}}(\mathbf{q},\omega)\text{,}
\label{eqn1}
\end{equation}%
where $\epsilon_{0}(=2.4)$ is the background dielectric constant. The external potential $V_{ll^{\prime}}(\mathbf{q})$ is expressed as $v_{q}e^{-q|l-l^{\prime}|I_{c}}$, where $v_{q}(=2\pi e^{2}/q)$ is the 2D bare Coulomb potential of a 2D electron gas, and the layer distance $I_{c}$ is chosen as $3.35{\AA}$ \cite{CRCPress;9781138556522}. The induced potential in the third term reveals the complex dynamic screening due to the intralayer and interlayer Coulomb interactions. The layer-dependent bare polarization function, being determined by energy bands and wave functions, is expressed as

\begin{equation}
\begin{array}{l}
P^{(1)}_{mm^{\prime}}(\mathbf{q},\omega)=2\sum\limits_{k}\sum\limits_{n,n^{\prime}}\sum\limits_{h,h^{\prime}=c,v}
\biggl(\sum\limits_{i}u^{h}_{nmi}(\mathbf{k})u^{h^{\prime}*}_{n^{\prime}m^{\prime}i}(\mathbf{k+q})\biggr)\\
\times\biggl(\sum\limits_{i^{\prime}}u^{h*}_{nm^{\prime}i^{\prime}}(\mathbf{k})
u^{h^{\prime}}_{n^{\prime}m^{\prime}i^{\prime}}
(\mathbf{k+q})\biggr)\times\frac{f(E^{h}_{n}(\mathbf{k}))-f(E^{h^{\prime}}_{n^{\prime}}(\mathbf{k+q}))}
{E^{h}_{n}(\mathbf{k})-E^{h^{\prime}}_{n^{\prime}}(\mathbf{k+q})
+\hbar\omega+i\Gamma}
\text{.}
\end{array}
\label{eqn2}
\end{equation}
In this notation, ${u^h_{nmi}}$ is the amplitude of the wave function on the $i$th sublattice of the $m$-th layer, arising from the valence/conduction state ($h=c$ and  $v$) of the $n$-th energy band. $f(E^{h}_{n}(\mathbf{k}))=1/\{1+\exp[(E^{h}_{n}(\mathbf{k})-\mu(T))/k_{B}T]\}$ is the Fermi-Dirac distribution function. $k_{B}$, $\mu(T)$ and $\Gamma$ stand for the Boltzmann constant, chemical potential and the energy width due to various deexcitation mechanisms, respectively. Moreover, the layer-dependent dielectric function is defined by the linear relationship between the effective and external potentials, i.e.,

\begin{equation}
\epsilon_{ll^{\prime}}(\mathbf{q},\omega)=\epsilon_{0}\delta_{ll^{\prime}}-
\sum\limits_{m}V_{lm}(\mathbf{q})P^{(1)}_{m,l^{\prime}}(\mathbf{q},\omega)\text{.}
\label{eqn3}
\end{equation}%

Making use of Eq.\ (\ref{eqn3}), one can express Eq. ((\ref{eqn1}) as a linear tensor equation. The effective potential tensor is the inverse of the dielectric function tensor multiplied by the external potential tensor. The dimensional energy loss function is useful in understanding the inelastic scattering probability of  EELS measurements \cite{PRB74;085406,Springer;2011}

\begin{equation}
\mathbf{Im}[-1/\epsilon]\equiv\sum\limits_{l}\mathbf{Im}\biggl[-V_{ll}^{eff}(\mathbf{q},\omega) \biggr]
/\biggl(\sum\limits_{l,l^{\prime}}V_{ll^{\prime}}(\mathbf{q})/3\biggl)\text{.}
\label{eqn4}
\end{equation}%
Equations ((\ref{eqn1}) through ((\ref{eqn4}), which include all the atomic and
Coulomb interactions, are applicable to any layered graphene system.

\section{Loss function and plasmon excitations}
\label{sec4}

Many types of single-particle excitation (SPE) channels and plasmon modes are now carefully investigated  in detail, especially for the strong frequency dependence of the electronic excitations on the magnitude of transferred momentum ($q$) and $E_F$. The predicted results could be verified by high-resolution  EELS \cite{PRB88;075433,Carbon114;70} and inelastic light scattering spectroscopy
\cite{Nature121;619,PRB81;205410}.

\medskip
\par

The unusual energy dispersions and van Hove singularities lead to the special structures in the  SPE spectra, There exist six independent bare polarization functions, namely,  ${P_{11}}$, ${P_{22}}$ ${P_{33}}$, ${P_{12}}$, ${P_{13}}$, and ${P_{23}}$. This number is larger than  that (four) in AAA (four), ABA and ABC stackings. For  pristine trilayer AAB stacking [${E_F=0}$ in Fig. 2(a)], three pairs of energy bands can create nine categories of interband transition channels, being expressed as ${i^v\rightarrow\,j^c}$ for $i$ and $j$=1, 2 ;3. The first pair of valence and conduction bands gives rise to two asymmetric peak structures in the lower-frequency polarization functions at small transferred momenta, e.g., ${P_{ij}}$ at ${\omega\,=0.01}$ and 0.12 eV's/${\omega\,=0.015}$ and 0.12 eV's for ${q=0.005}$/${q=0.01}$ ${1/\AA}$ (the black/red curves in Figs. 4 and 5). The real and imaginary parts, respectively, show the square root divergent peak and a pair asymmetric peak structures, directly reflecting the Kramer-Kronig relations. It should be noticed that the Coulomb excitations belong to the non-vertical transitions. However, the valence to conduction band-edge state transitions could survive under small $q$'s. With  increasing excitation frequency,  the special structures, without the specific form, occur in the range of 0.2 eV${<\omega\,<}$ 0.4 eV. The main reason that such composite structures are associated with the large joint density-of-states from the band-edge states of the first oscillatory valence band and the second sombrero-shaped conduction band (the second sombrero-shaped valence band and the first oscillatory conduction band). The second pair of shallow sombrero-shaped energy bands only induce a square root peak (a pair of asymmetric structures) in Im ${[P_{ij}]}$'s Re\ (${[P_{ij}]}$s within the range of 0.4 eV ${<\omega\,<}$ 0.6 eV. The second (third) valence states could also be excited to the third (second) conduction bands, where the band-edge states near the K point might lead to the special structures at 0.6 eV ${<\omega\,<}$ 0.8 eV. Finally, the third pairs of parabolic bands show  shoulder structure in Im\ ${[P_{ij}]}$'s/the logarithmically symmetric peak in Im\ ${[P_{ij}]}$'s.

\medskip
\par

The energy loss spectra of  undoped trilayer AAB stacking can exhibit a very prominent peak, clearly illustrating the dominating band structure effect. The plasmon frequency lies in the range of 0.145 eV$-$0.22 eV for the momentum range of ${q=0.005-0.02}$ ${1/\AA}$, as shown in Fig. 6(a) by the distinct solid curves. Its intensity and frequency, respectively, decreases and grows as $q$ is increased. This strong plasmon mode is related to the interband transitions of the first pair of oscillatory energy bands [Fig. 2(a), Figs. 3-4], in which it directly reflects the significant van Hove singularities nearest to the Fermi level. The low frequency plasmon peak is prominent, since it is far away from the Landau damping due to the higher frequency interband excitations of ${1^2\rightarrow\,2^c}$ and ${2^v\rightarrow\,1^c}$. Also, other low intensity plasmon peaks could be observed in the screened response functions, e.g., the ${0.32-0.38}$ eV peaks arising from the first/second valence band and the second/first conduction band [the inset of Fig. 6(a)]. Specifically, the low frequency collective excitations, with the significant peaks, are revealed in narrow gap carbon nanotubes. Free electrons and holes cannot survive in these two systems, while the creation of the low frequency prominent plasmon modes is attributed to the 2D oscillatory bands/the 1D parabolic bands with the large DOS; the dimension- and wave function-dependent strong Coulomb interactions in the long wavelength limit.

\medskip
\par

Electron doping can drastically modify the collective excitations and the special features in the single-particle response functions, as clearly demonstrated in Figs. 4 and 5. For example, as for the low-doping case when ${E_F=0.1}$ eV (the blue curves), certain intraband and interband excitations are, respectively, allowed or  forbidden by the Pauli exclusion principle. The low frequency interband ${1^v\rightarrow\,1^c}$ channel is replaced by the intraband ${1^c\rightarrow\,1^c}$, so that the only special structure at ${\omega\sim\,0.022}$ meV mainly comes from the  Fermi momentum states, but not the band edge ones. The linear energy dispersions near $E_F$ also lead to the asymmetric square root peak \cite{PRB34;979}. In the range of 0.2 eV $<\omega<$ 0.4 eV, the special features are due to the extra ${1^c\rightarrow\,2^c}$  excitations and the original interband ${1^v\rightarrow\,2^c}$ excitations simultaneously, in which they are related to the band edge states. With the further increase of frequency, the identical structure at ${\omega\sim\,0.522}$ eV is induced by the same interband ${2^v\rightarrow\,2^c}$. The ${3^v\rightarrow\,1^c}$ channel is changed into the ${1^c\rightarrow\,3^c}$ one at higher frequencies, and the other is similar, e.g., the ${1^v\rightarrow\,3^c}$, ${2^v\rightarrow\,3^c}$ and ${3^v\rightarrow\,3^c}$ channels.  The variation of the Fermi level will modify the  SPE's which even results in the creation or destruction of the different plasmon modes.  We also notice that the threshold asymmetric structure at the lowest frequency could survive for any chosen Fermi level, and so does the very prominent acoustic plasmon peak.

\medskip
\par

After the electron doping, the screened response functions, revealing the distinct collective excitations and the various Landau dampings, present the rich and unique phenomena [Figs. 6(a) through 6(c) and 7(a) through 7(c)]. For the lower doping of ${E_F=0.1}$ [Fig. 6(b)], the frequency and intensity of the first plasmon peak are slightly reduced under the same transferred momenta, compared with those of ${E_F=0}$ [Fig. 6(a)]. The critical mechanisms are in sharp contrast with each other, in which the former and latter cases, respectively, correspond to the Fermi momentum states and the valence band edge states. That is, the collective oscillations of charge carriers mainly come from the free conduction electrons and the valence ones in the first pair of oscillatory energy bands. The plasmon is very sensitive to the direction of the transferred momentum, e.g., the great reduction in the strength and frequency of the low frequency plasmon  as a result of the significant interband Landau dampings [Fig. 6(c)]. Furthermore, the observable plasmons at higher frequencies, which are located in the range of 0.2 eV$<\omega<$0.4 eV [the insets in Figs. 6(b) and 6(c)] are two distinct modes due to the $1^v\rightarrow\,2^c$ and $1^c \rightarrow\,2^c$  channels. However, there is one interband plasmon peak for the undoped case [the inset in Fig. 6(a)] because of the approximately same excitation frequency of the $1^v\rightarrow\,2^c$ and $2^v \rightarrow\,1^c$ channels.  With increased Fermi energy [e.g., ${E_F}$=0.3 eV in Fig. 7(a)], there are more occupied states (free electrons) arising from the first and second conduction bands, so that two intraband excitation channels, the $1^c\rightarrow\,1^c$ and $2^c \rightarrow\,2^c$ ones,  would enhance the bare and screened response functions. Consequently, the first plasmon mode has an enhanced frequency and strength. Similar excitation phenomena appear at the higher-$E_{F}$ cases, such as, ${E_F=0.5}$ eV and 0.7 eV [Figs. 7(b) and 7(c)] crossing the third conduction band [Fig. 2(a)]. Also, the energy loss spectra clearly display  drastic changes in the number, frequency and intensity of the higher frequency plasmon peaks for variations of $E_F$ [the insets in Fig. 6(a) through 6(c) and 7(a) through 7(c)].

\medskip
\par

Clearly, the (momentum-frequency) phase diagrams are diversified by increasing the  doping density, as  indicated in Fig. 8(a) through 8(f). The single-particle regions are enriched by doping. For the pristine system at zero temperature, the nine available interband excitations yield well-defined boundaries [Fig. 8(a)], being characterized by the band edge states [Fig. 2], but not the Fermi momentum states. Generally, the whole phase space is almost full of the electron-hole excitations. Furthermore, the specific region of the  small  $q$$^{,}$s and $\omega$$^{,}$s exhibit very weak electron-hole excitations [the left-hand down corner in Fig. 8(a)]. Similar behavior is revealed in the doped case, without  the Landau dampings in the specific region [the crossed regions under ${E_F\neq\, 0}$ in Figs. 8(b) through 8(f)]. The original boundaries are drastically modified by the Fermi level if the final Coulomb scattering states are associated with the partially occupied conduction energy bands, e.g., $E_F$=0.1 eV [Fig. 8(b)], 0.3 eV [Figs. 8(c) and 8(d)], 0.5 eV [Fig. 8(e)], and 0.7 eV [Fig. 8(f)]. The interband valence-state excitation regions are greatly reduced by the increasing $E_F$, being more suitable for the existence of the low frequency acoustic plasmon [Fig. 8(f)]. Furthermore, the new excitation boundaries are created by the doping effects. For example, three kinds of extra electron-hole boundaries principally come from the ${1^c\rightarrow\,1^c}$, ${1^c\rightarrow\,1^c}$ $\&$ ${1^c\rightarrow\,1^c}$ channels if the Fermi level is between the first and  second conduction energy bands, such as the SPE regions under ${E_F=0.1}$ eV by the red notations.

\medskip
\par

\medskip
\par

There exists one acoustic plasmon for any doping,  with its frequency approximately behaving as ${\sqrt q}$  in the long wavelength limit. In  the pristine system, this mode is created by the valence electrons associated with a narrow energy gap, whereas it comes from all the conduction electrons related to the intraband excitations. The low frequency plasmon in trilayer AAB stacking, which resembles the acoustic mode for the 2D electron gas \cite{Springer;2011} further illustrates the critical mechanism due to the large valence DOS of the oscillatory band or the free conduction electrons [Fig. 2(a)]. However, the critical momentum ($q_c$) strongly depends on the Fermi level, since the higher energy interband excitations are mainly determined by it. The low frequency plasmon would disappear [Figs. 8(a) through 8(d)] or merge with the optical plasmon mode [Figs. 8(e) and 8(f)] after entering  the interband electron-hole excitation region for ${q>q_c}$. A simple relation between $q_c$ and $E_F$ is not available because of the complicated non-vertical interband Coulomb excitations, such as, $q_c$=0.054 ${1/\AA}$, 0.03 ${1/\AA}$, 0.02 ${1/\AA}$, 0.04 ${1/\AA}$ and $\sim$0.1 ${1/\AA}$, respectively, corresponding to ${E_F=0}$, 0.1 eV, 0.3 eV, 0.5 eV and  0.7 eV. This plasmon  mode is very prominent for the pristine and high-$E_F$ cases [Figs. 8(a) and 8(f)]; i.e., it is relatively easy to measure the low frequency collective excitations in the absence (presence) of doping (high doping). As for the higher frequency plasmons, the energy loss spectra may yield one or two undamped modes,  sensitive to the Fermi level. One or two optical modes, respectively, correspond to [$E_F$=0; 0.3 eV] and [$E_F$-0.1 eV, 0.5 eV; 0.7 eV]. In addition, the strongly hybridized optical plasmon modes might survive for a certain range of the Fermi level, e.g., those within ${E_F\sim\,0.3-0.5}$ eV [Figs. 8(c) through 8(e)]. The optical plasmon in the pristine system is due to the interband ${1^v\rightarrow\,2^c}$ and ${1^v\rightarrow\,2^c}$ excitations. On the other hand, the diverse critical mechanisms, being sensitive to $E_F$, are revealed for two distinct optical modes for the doped cases. For example, when ${E_F=0.1}$ [$E_F$=0.7 eV], the second and third plasmons, with frequencies higher than 0.2 eV and 0.3 eV [0.65 eV and 0.95 eV], are dominated by the interband ${1^c\rightarrow\,2^c}$ and  ${1^v\rightarrow\,2^c}$ excitations, as for the ${1^v\rightarrow\,3^c}$] excitations at small transferred momenta. In general, two optical plamsons are closely related to the multi-interband excitation channels [Figs. 8(b)-8(f)].

\medskip
\par
\medskip
\par

The significant differences encompassed by the distinct trilayer stakcing arrangements are worthy of a detailed comparison. Clearly, the AAB, ABC, AAA and BAA stackings exhibit rich and unique Coulomb excitations, certainly illustrating the geometry-diversified phenomena. From the calculated bare and screened response functions \cite{PRB74;085406},  the geometric symmetries, which principally determine the various interlayer hopping integrals, fully dominate the low lying energy bands and thus the single-particle and collective excitations in the frequency range below ${\sim\,1}$ eV. Evidently, the electron-hole excitation boundaries/regions are related to the band-edge states with the van Hove singularities and the Fermi-momentum states, being sensitive to the stacking-generated energy dispersions. As a result, they are quite different among four kinds of stacking configurations. Concerning pristine AAB and ABC stackings, the first pair of valence and conduction bands nearest to the Fermi level [the two pairs of Dirac cones below and above the Fermi level, but not the specific one through $E_F$] can create a very strong special structure in the bare polarizability and a prominent peak in the energy loss spectrum at low frequency. That is, the low frequency acoustic plasmon is identified as coming from the valence states with the large DOS in a narrow-gap (AAB) or gapless (ABC) system, or the free electrons and holes (AAA) due to the strong interlayer atomic interactions. On the other hand, it is very challenging to observe this plasmon in the ABA stacking, mainly owing to the low DOS valence band-edge states \cite{arXiv:1803;10715}.   Certainly, such low frequency plasmons can survive in any doped graphene, since it is induced by all the free carriers. There exist dramatic transformations in the available low frequency excitation channels as the Fermi energy is gradually increased from ${E_F=0}$. In general, a simple relation between the main features of the acoustic plasmon and $E_F$ is missing for low doping except for  ABA stacking. That is, the frequency and intensity do  present an analytic  dependence on $E_F$. Such result is attributed to strong competition in AAB and ABC stackings between  the low frequency interband excitations related to the valence band edge states and the intraband excitations associated with the conduction electrons [the opposite variations of free electrons and holes in AAA stacking]. However, the reverse is true for the ABA stacking. Regarding the higher frequency collective excitations, all the pristine  AAB, ABA and ABC stackings show one detectable optical mode arising from the most significant interband excitations. No optical plasmons are obtained in the undoped ABA stacking. Specifically, there are one or two optical modes in any doped system, being enriched by a lot of interband excitations due to the valence and conduction electrons.  Many kinds of SPE channels and plasmon modes have been explored in detail, especially  the strong dependence of the electronic excitations on the magnitude of the transferred momentum ($q$) and $E_F$. It should be possible for the predicted results to be verified by high-resolution EELS \cite{PRB88;075433,Carbon114;70} and inelastic light scattering spectroscopy \cite{hermanson,Ohta,Jin,Nature121;619,PRB81;205410}.

\section{Concluding remarks}
\label{sec5}

We have demonstrated that  trilayer AAB stacking of graphene exhibit unique electronic properties leading to diverse Coulomb excitations. The lower stacking symmetry leads to three pairs of unusual energy dispersions, i.e., the oscillatory, sombrero-shaped, and parabolic bands. The first two possess large and special van Hove singularities, especially for the first pair nearest to the Fermi level. Consequently, for  pristine systems, there are nine categories of valence$\rightarrow$conduction band transitions. The special structures in the bare response functions cover the square root asymmetric peaks and the shoulder structures [the pairs of anti-symmetric prominent peaks and logarithmically symmetric peaks] in the imaginary [real] part. The threshod channel, ${s_1^c\rightarrow\,s_1^c}$, can create significant single-particle excitations and the strong collective excitations. The low frequency acoustic plasmon, being characterized by the pronounced peak in the energy loss spectrum, is purely due to the large DOS in the oscillatory valence and conduction bands and the narrow energy gap.  Furthermore, its intensity and frequency could be reduced at finite temperatures. Similar plasmon modes can be excited in a narrow gap carbon nanotube. The critical mechanism for the creation of this plasmon is thoroughly transformed into all the intraband conduction band excitations, for which the effective channels and the critical transferred momenta strongly depend on the Fermi level.

\medskip
\par
After electron/hole doping, the interband particle-hole excitation regions are drastically modified and the extra intraband ones are generated by varying $E_F$. Moreover, one or two higher frequency optical plasmon modes survive for various Fermi energies. They are closely related to the specific excitation channels or the strongly overlapped multi-channels, being sensitive to the Fermi level and transferred momenta. There are certain important differences among the trilayer AAB, ABC, ABA and AAA stackings, such as the boundaries of the various intraband and interband electron-hole excitations, and the mechanism, number, strength, frequency and mode of the collective excitations. To fully explore the geometry-enriched Coulomb excitations, the above-mentioned theoretical predictions require  experimental verifications.

\par\noindent {\bf Acknowledgments}
This work was supported in part by the National Science Council of Taiwan, the Republic of China, under Grant Nos. NSC 105-2112-M-006 -002 -MY3.

\newpage
\renewcommand{\baselinestretch}{0.2}

\newpage
\par

\begin{figure}[t]
\centering
\includegraphics[width=3.4in,height=3.1in,keepaspectratio]{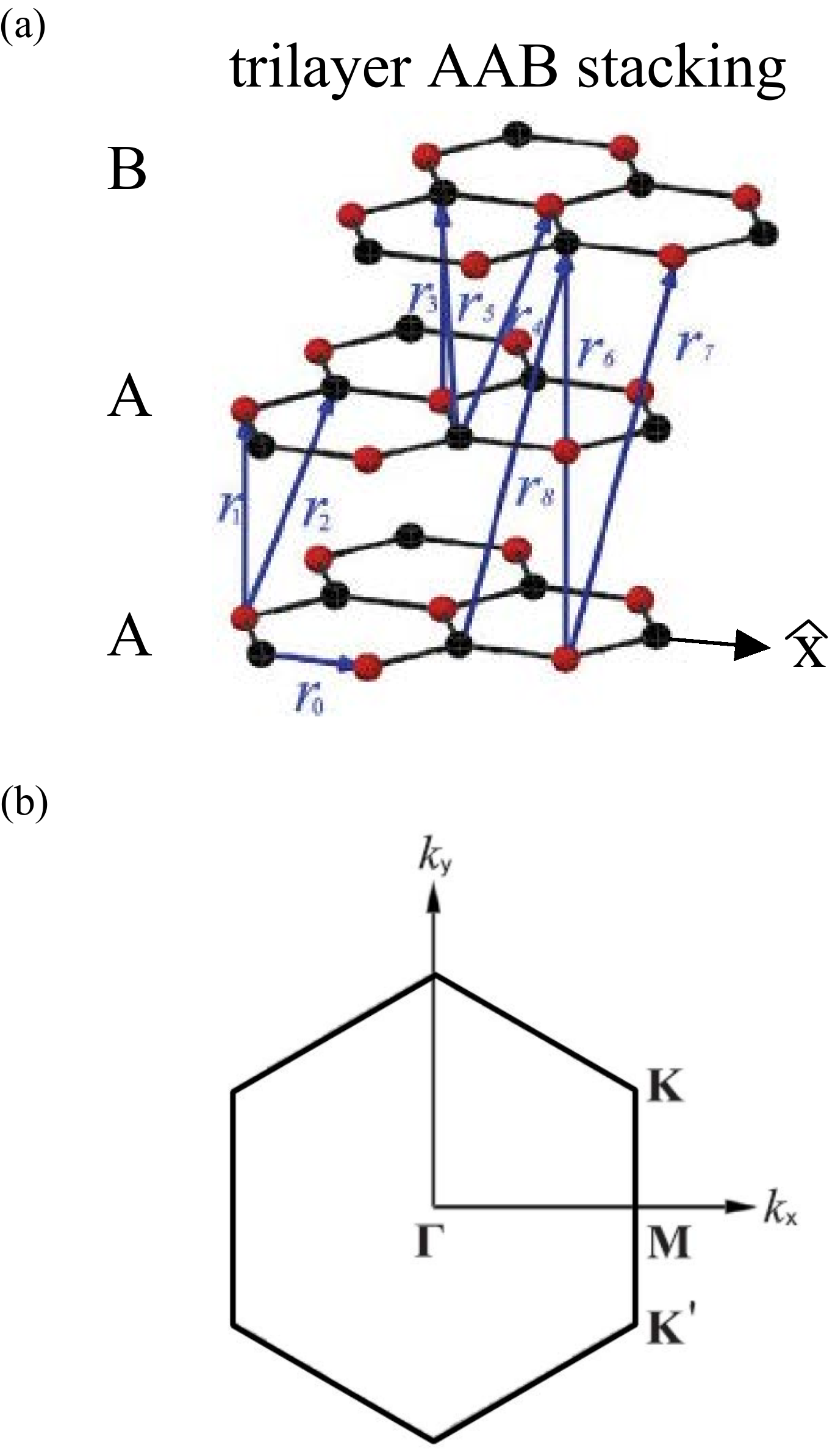}
\caption{(color online) The geometric structure of the trilayer AAB-stacked graphene in the
presence of various intralayer and interlayer hopping integrals and its first Brillouin zone.}
\label{FIG:1}
\end{figure}

\newpage
\par

\begin{figure}[t]
\centering
\includegraphics[width=3.4in,height=3.1in,keepaspectratio]{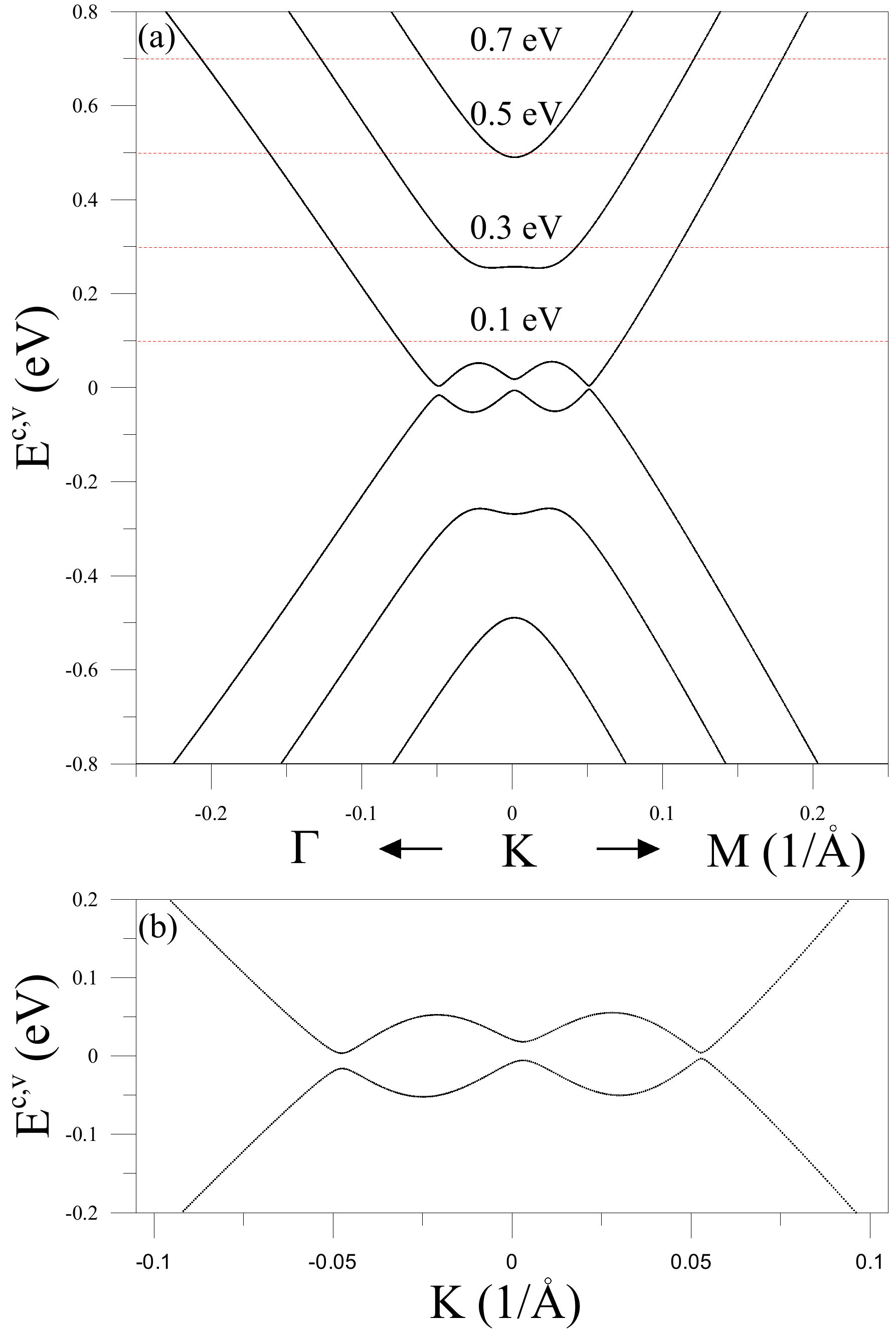}
\caption{(color online) (a) The low-lying valence and conduction bands for trilayer AAB stacking,
 with (b) a narrow gap of ${E_g\sim\,8}$ meV arising from oscillatory energy bands. Also shown by the
dashed curves are those from the first-principles method.
}
\label{FIG:2}
\end{figure}

\newpage
\par

\begin{figure}[t]
\centering
\includegraphics[width=3.4in,height=4.1in,keepaspectratio]{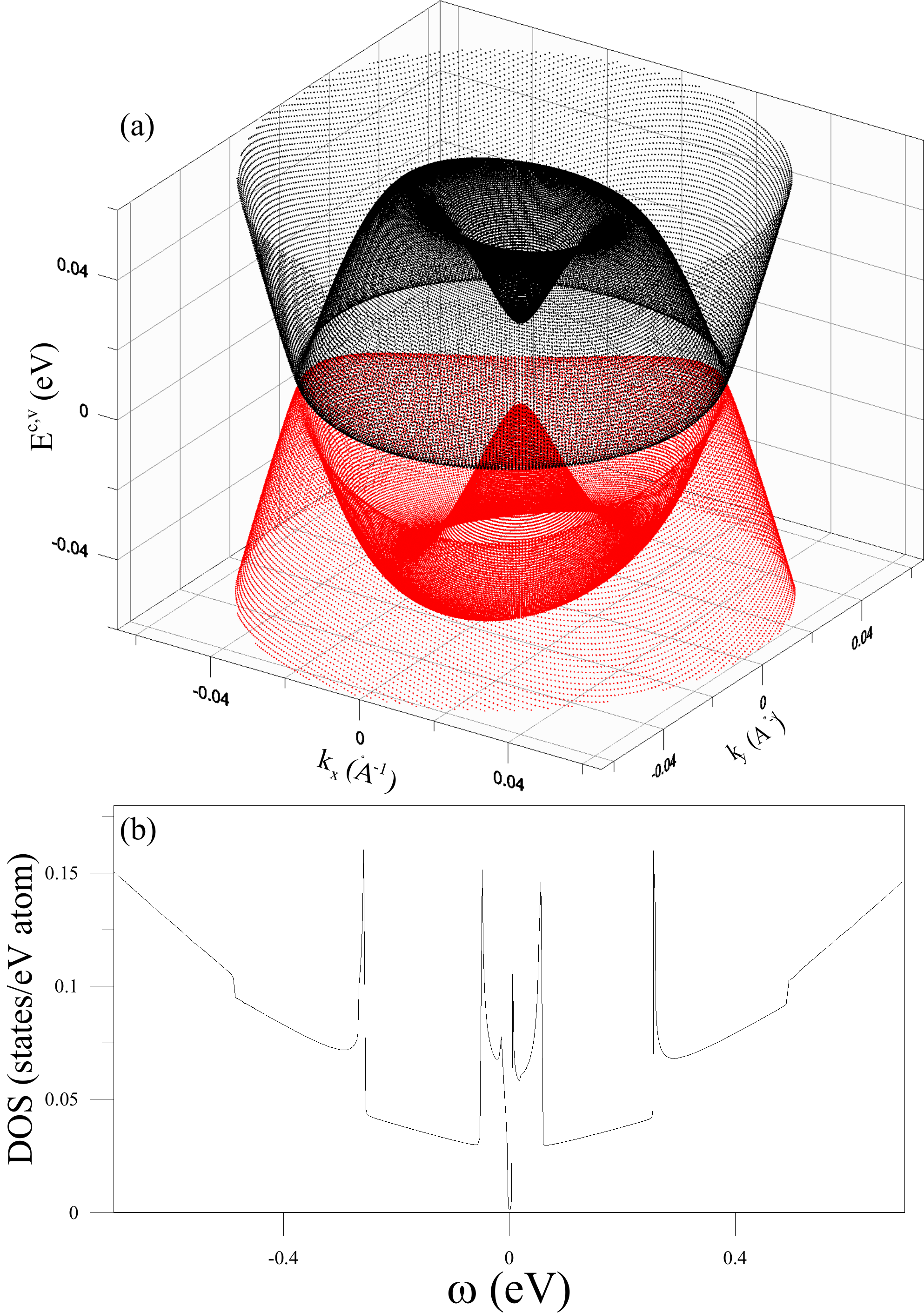}
\caption{(color online) (a) The 3D energy bands near the Fermi level and (b)
the low energy density-of-states.
}
\label{FIG:3}
\end{figure}

\newpage
\par

\begin{figure}[t]
\centering
\includegraphics[width=3.4in,height=4.1in,keepaspectratio]{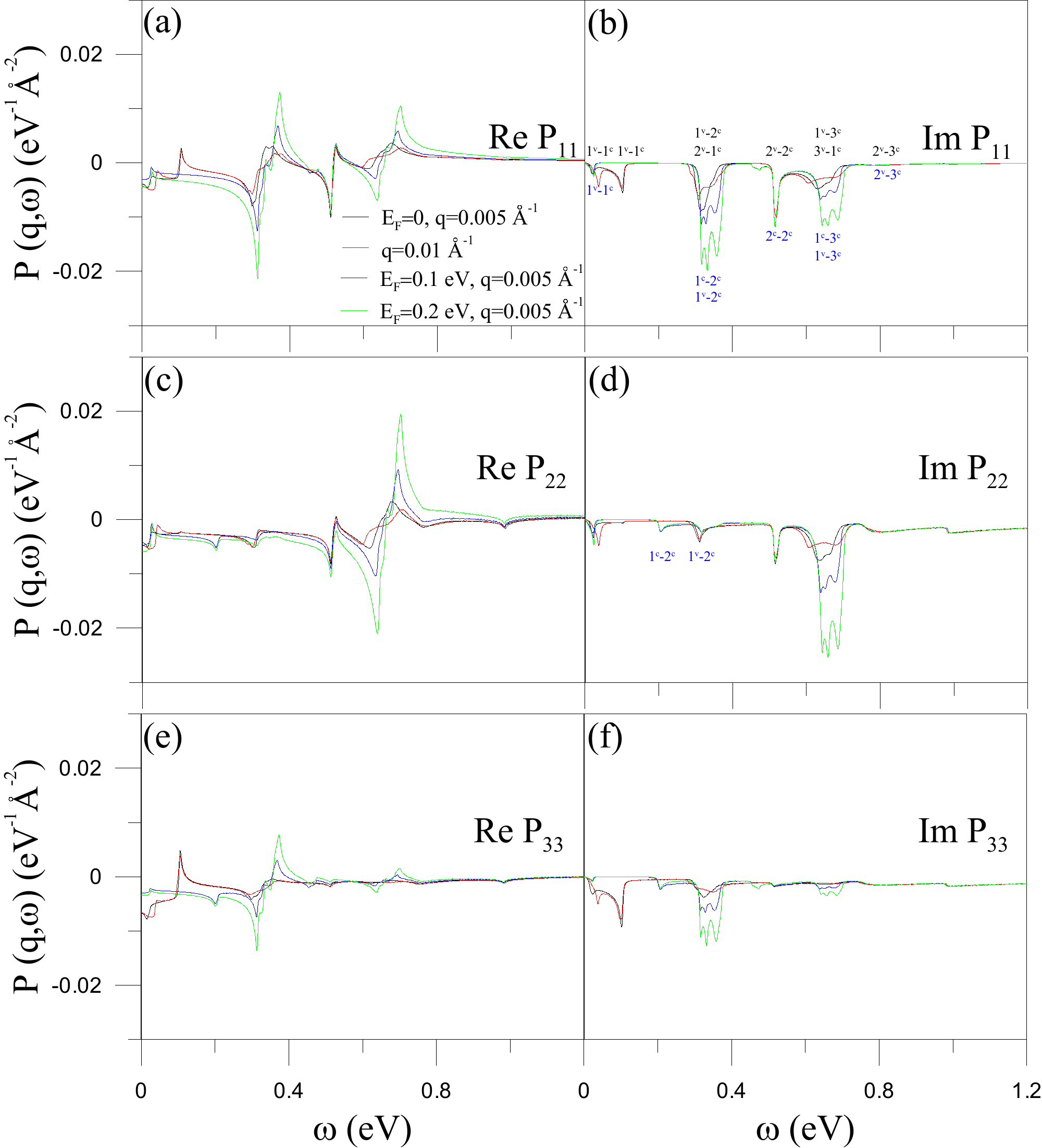}
\caption{(color online) The bare polarization functions for trilayer AAB
stacking graphene for chosen Fermi energies and transferred momenta:
(a) Re ${[P_{11}]}$, (b) Im [${P_{11}]}$, (c) Re ${[P_{22}]}$, (d) Im ${[P_{22}]}$, (e) Re ${[P_{33}]}$, and (f) Im ${[P_{33}]}$. }
\label{FIG:4}
\end{figure}

\newpage
\par
\begin{figure}[t]
\centering
\includegraphics[width=3.4in,height=4.1in,keepaspectratio]{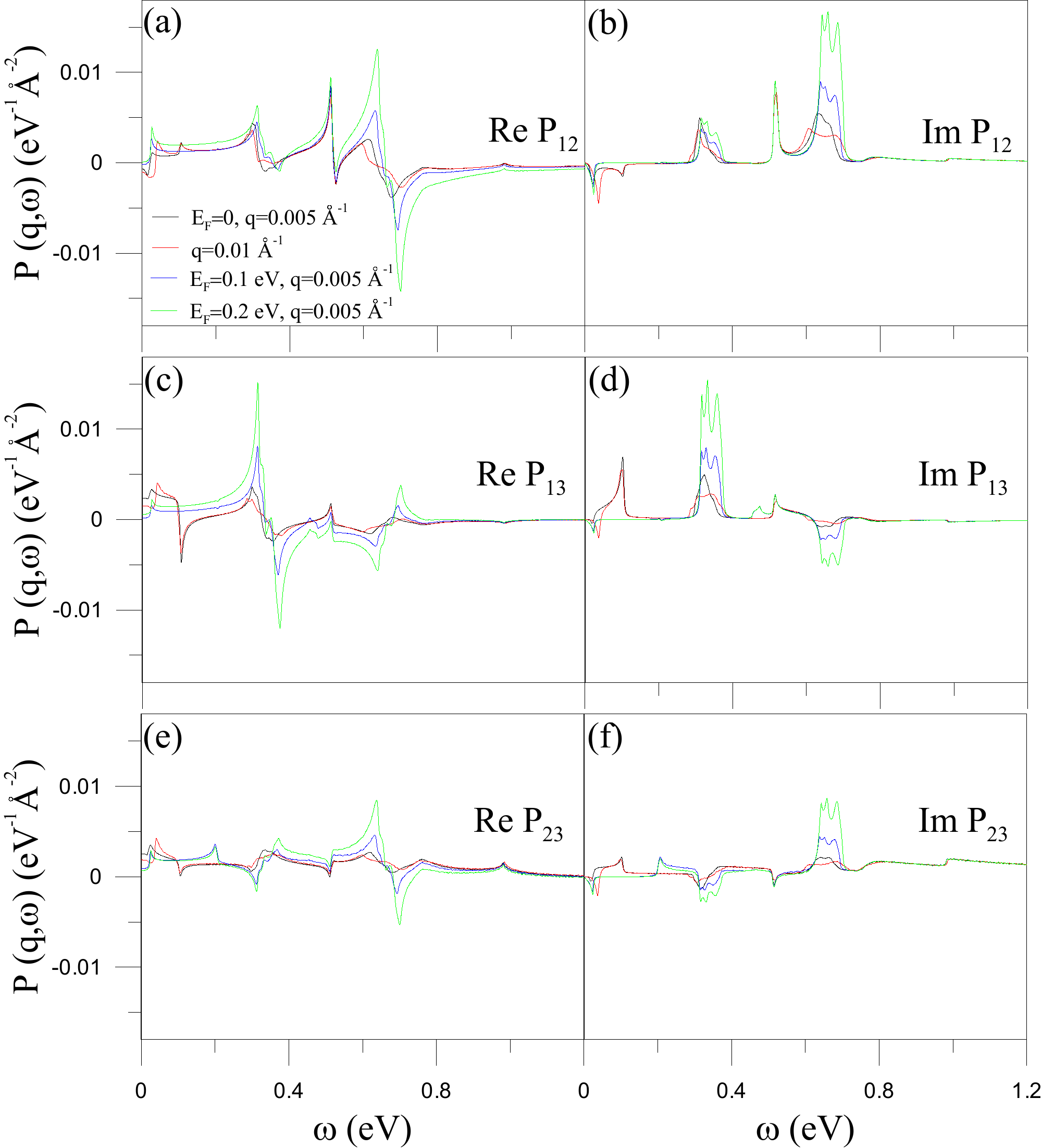}
\caption{(color online) Plots similar to Figs. \ \ref{FIG:4}(a) through
\ \ref{FIG:4}(f), shown for: (a) Re ${[P_{12}]}$, (b) Im ${[P_{12}]}$,
(c) Re ${[P_{13}]}$, (d) Im ${[P_{13}]}$,
(e) Re ${[P_{23}]}$, and (f) Im ${[P_{23}]}$.}
\label{FIG:5}
\end{figure}

\newpage
\par
\begin{figure}[t]
\centering
\includegraphics[width=3.4in,height=4.1in,keepaspectratio]{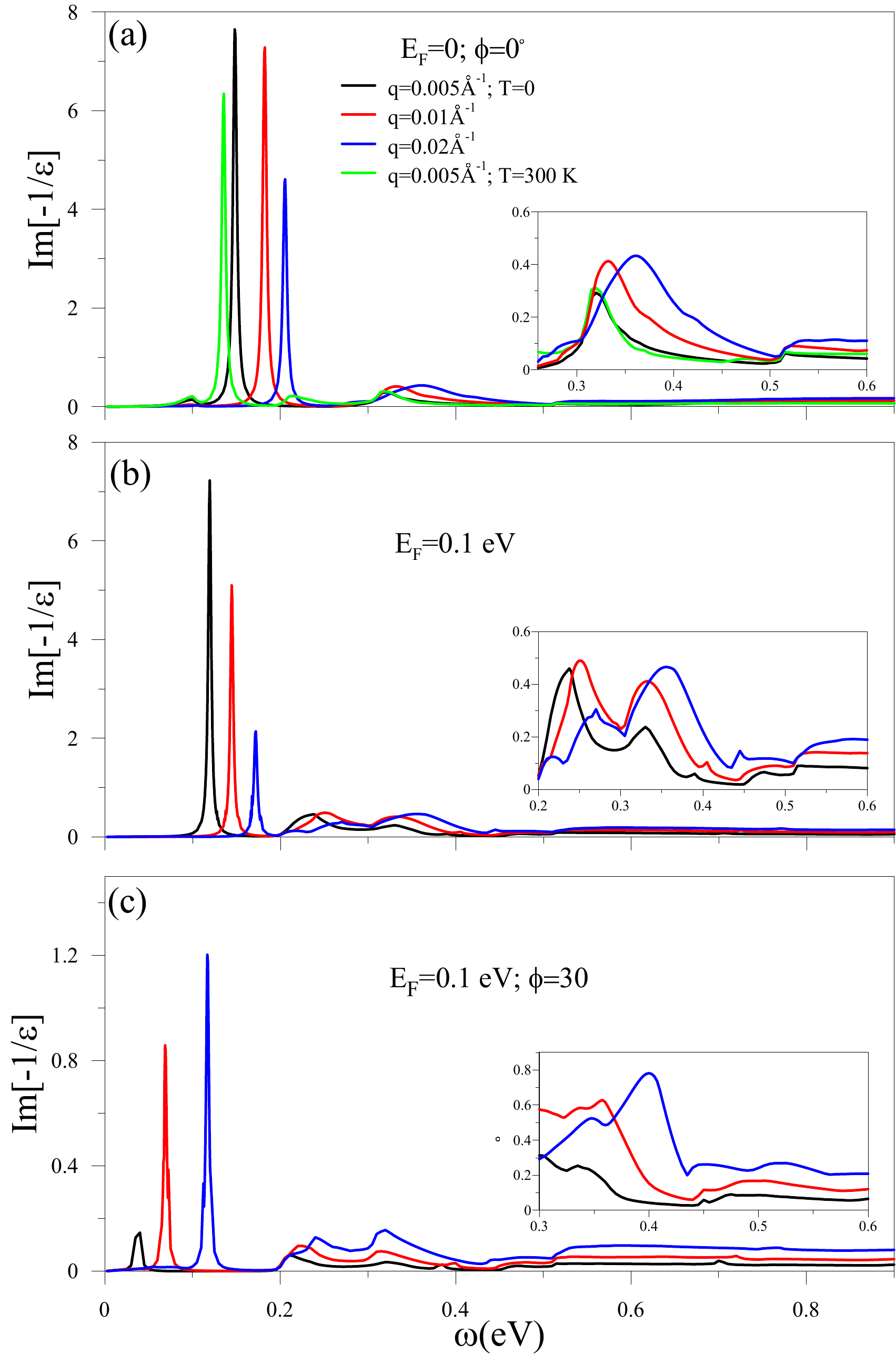}
\caption{(color online) The energy loss functions for the chosen
transferred momenta and direction  ${\phi\,=0^\circ}$ when
(a) ${E_F=0}$, and (b) 0.1 eV. Also shown in (c) are those
at ${phi\,=30^\circ}$ and ${E_F=0.1}$ eV. The insets show the
higher frequency loss spectra covering one or two optical plasmon modes.}
\label{FIG:6}
\end{figure}

\newpage
\par
\begin{figure}[t]
\centering
\includegraphics[width=3.4in,height=3.1in,keepaspectratio]{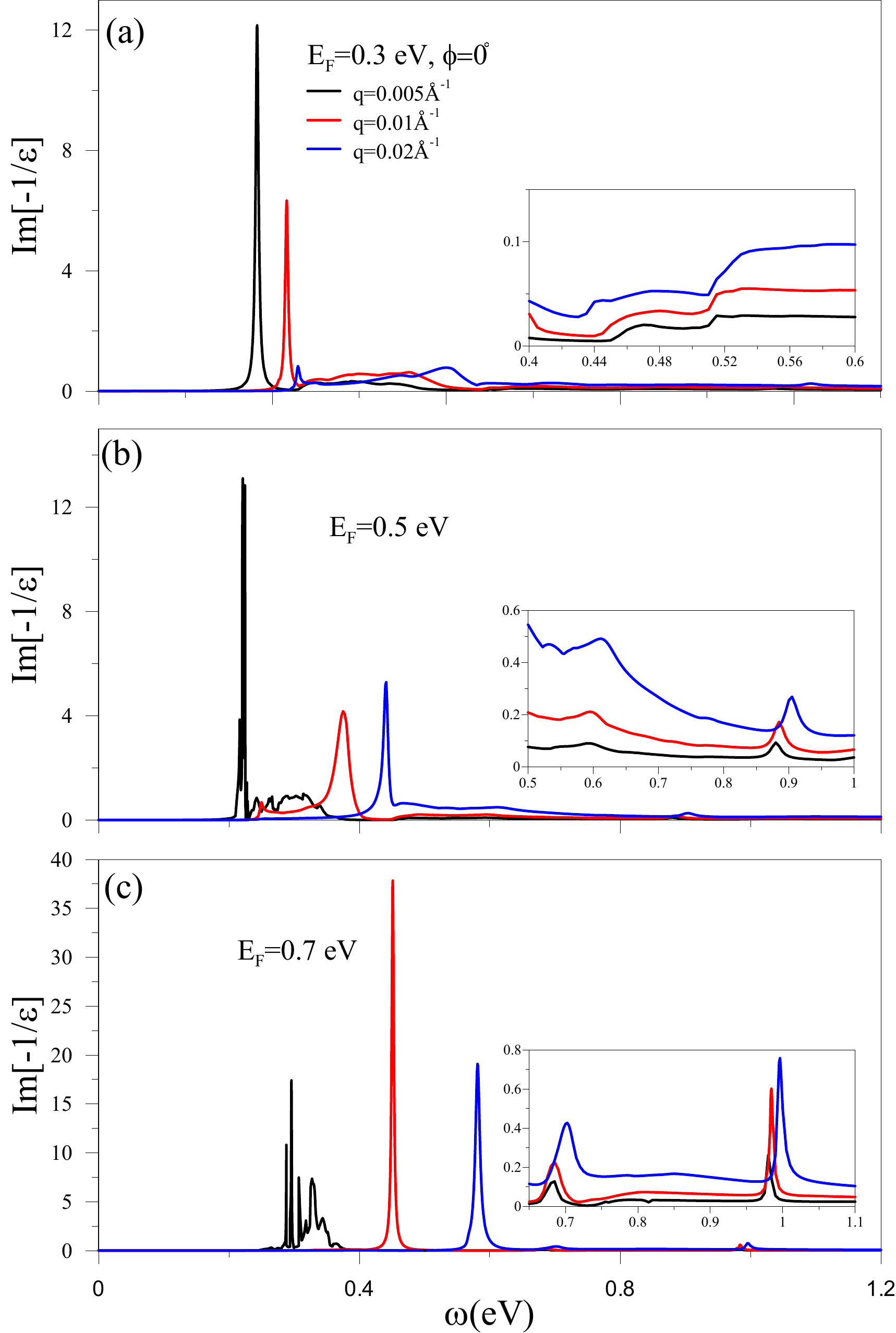}
\caption{(color online)  The energy loss functions under the various transferred momenta and ${\phi\,=0^\circ}$ for (a) ${E_F=0.3}$. (b) 0.5 eV, and (c) 0.7 eV. The insets show the higher-frequency loss spectra covering one or two optical plasmon modes.
}
\label{FIG:7}
\end{figure}

\newpage
\par
\begin{figure}[t]
\centering
\includegraphics[width=3.4in,height=3.1in,keepaspectratio]{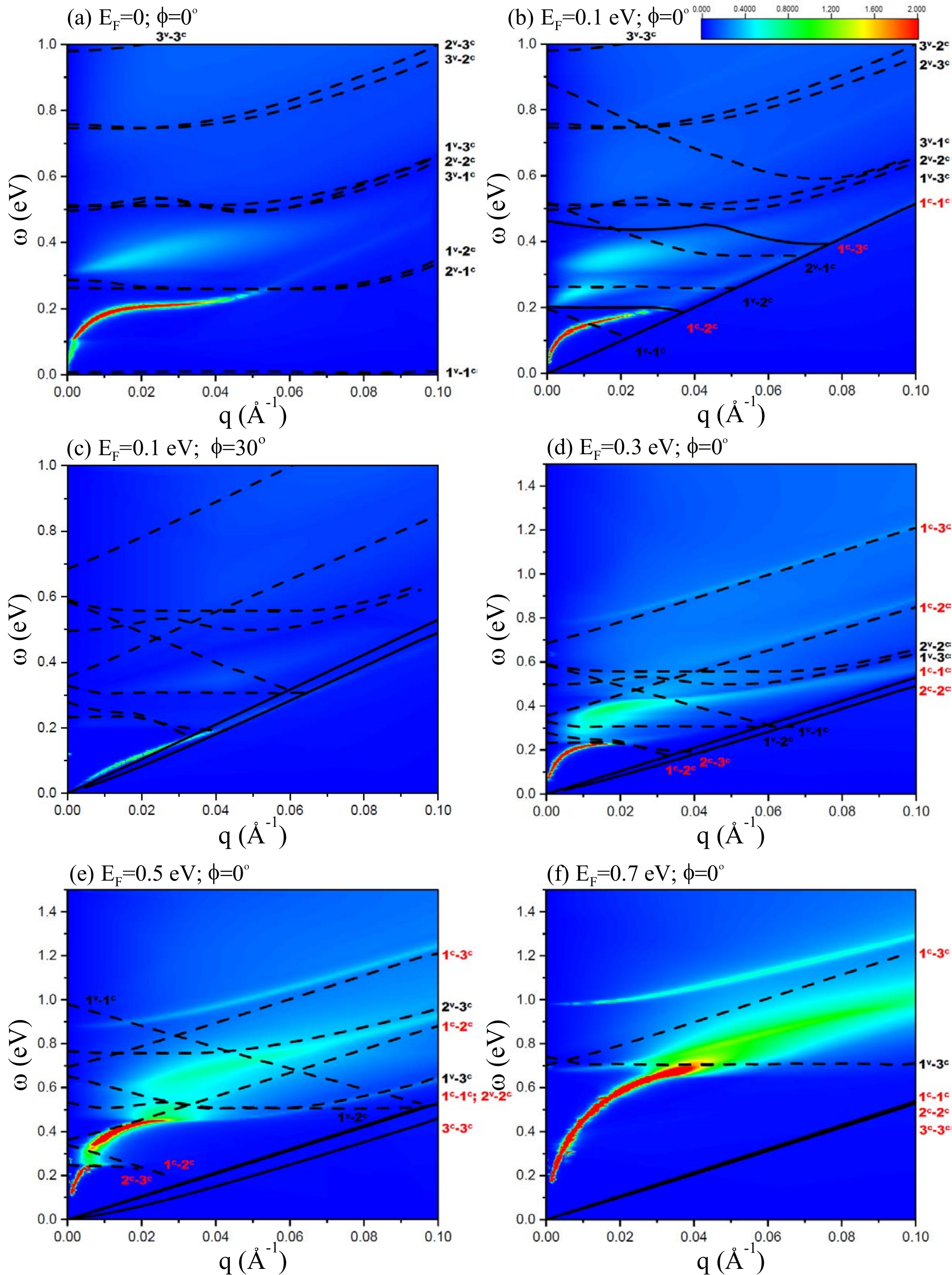}
\caption{(color online)  The (momentum, frequency)-phase diagrams of the trilayer AAB stacking under the various Fermi levels: (b) ${[E_F=0, \phi\,=0^\circ]}$, (b) ${[E_F=0.1}$ eV, ${\phi\,=0^\circ]}$,
(c) ${[E_F=0.1}$ eV, ${\phi\,=30^\circ]}$, (d) ${[E_F=0.3}$ eV, ${\phi\,=0^\circ]}$,
(e) ${[E_F=0.5}$ eV, ${\phi\,=0^\circ]}$, and (f) ${[E_F=0.7}$ eV, ${\phi\,=0^\circ]}$. The region without the e-h excitations indicated by the crosses.
}
\label{FIG:8}
\end{figure}

\end{document}